\documentclass[10pt,conference]{IEEEtran}

\usepackage{amsmath}
\usepackage{cite}
\usepackage{graphicx}
\usepackage{amsfonts}
\usepackage{latexsym}
\usepackage{subfigure}
\usepackage{algorithm}
\usepackage{algorithmic}
\usepackage{color}
\usepackage{times}
\usepackage{epsfig, graphics}
\usepackage{bm}
\usepackage{subfigure}
\usepackage{graphicx,wrapfig,lipsum}
\usepackage{amsthm}

\begin{document}

\newtheorem{theorem}{Theorem}
\newtheorem{lemma}{Lemma}
\newtheorem{conjecture}{Conjecture}
\newtheorem{corollary}{Corollary}
\newtheorem{definition}{Definition}
\newtheorem{scheme}{Scheme}
\newcommand{\argmax}{\arg\!\max}
\newcommand{\rev}[1]{{\color{red}#1}} 
\newcommand{\pound}{\operatornamewithlimits{\gtrless}}
\IEEEoverridecommandlockouts

\title{Generative Adversarial Network in the Air: Deep Adversarial Learning for Wireless Signal Spoofing}

\author{\IEEEauthorblockN{Yi Shi, Kemal Davaslioglu, and Yalin E. Sagduyu}
\thanks{Yi Shi is with Virginia Tech, Blacksburg, VA, USA; Email: yshi@vt.edu. Kemal Davaslioglu is with Intelligent Automation, Inc., Rockville, MD, USA; Email: kdavaslioglu@i-a-i.com. Yalin Sagduyu is with Intelligent Automation, Inc., Rockville, MD, USA; Email: ysagduyu@i-a-i.com.}
\thanks{This effort is supported by the U.S. Army Research Office under contract W911NF-17-C-0090. The content of the information does not necessarily reflect the position or the policy of the U.S. Government, and no official endorsement should be inferred.}
\thanks{A preliminary version of the material in this paper was partially presented at ACM Conference on Security and Privacy in Wireless and Mobile Networks (WiSec) Workshop on Wireless Security and Machine Learning (WiseML), 2019. \cite{Shi19:wiseml}.}
\thanks{\copyright~2020 IEEE. Personal use of this material is permitted. Permission from IEEE must be obtained for all other uses, in any current or future media, including reprinting/republishing this material for advertising or promotional purposes, creating new collective works, for resale or redistribution to servers or lists, or reuse of any copyrighted component of this work in other works.}
}

\maketitle
\begin{abstract}
The spoofing attack is critical to bypass physical-layer signal authentication. This paper presents a deep learning-based spoofing attack to generate synthetic wireless signals that cannot be statistically distinguished from intended transmissions. The adversary is modeled as a pair of a transmitter and a receiver that build the generator and discriminator of the generative adversarial network, respectively, by playing a minimax game over the air. The adversary transmitter trains a deep neural network to generate the best spoofing signals and fool the best defense trained as another deep neural network at the adversary receiver. Each node (defender or adversary) may have multiple transmitter or receiver antennas. Signals are spoofed by jointly capturing waveform, channel, and radio hardware effects that are inherent to wireless signals under attack. Compared with spoofing attacks using random or replayed signals, the proposed attack increases the probability of misclassifying spoofing signals as intended signals for different network topology and mobility patterns. The adversary transmitter can increase the spoofing attack success by using multiple antennas, while the attack success decreases when the defender receiver uses multiple antennas. For practical deployment, the attack implementation on embedded platforms demonstrates the low latency of generating or classifying spoofing signals.
\end{abstract}


\section{Introduction}
By exploiting the open and shared nature of wireless spectrum, an  adversary can launch a \emph{spoofing attack} by mimicking transmissions from a legitimate user at the physical layer. The spoofing attack can be used for various adversarial purposes such as emulating primary users in cognitive radio networks and fooling signal authentication systems to intrude protected wireless networks. While wireless signals can be spoofed by recording a legitimate user's signals and replaying them later, such \emph{replay attacks} cannot necessarily capture all waveform, channel and device effects, and cannot provide an algorithmic mechanism to train itself against an authentication system, such as the one that uses a machine learning classifier to authenticate signals by analyzing the physical layer features in wireless signals and identifying the spoofed signals.

Compared to conventional feature-based machine learning techniques, \emph{deep learning} can model and represent high-dimensional spectrum dynamics by processing raw spectrum data without handcrafted feature extraction. Deep learning has been successfully applied to many applications in wireless communications such as spectrum sensing \cite{Kemal2018} and modulation recognition \cite{OShea2016}. An adversary can also apply deep learning and carefully design inputs to manipulate the behavior of a legitimate system in test or training time. Such attacks on machine learning have been studied under \emph{adversarial machine learning} \cite{HST17, AMLbook,Sec2} for various data domains such as computer vision and natural language processing (NLP).

There is also a growing interest in applying adversarial machine learning to the wireless domain. A deep learning-based jammer was studied in \cite{Yi18:Jamming, Tugba2018} to learn the transmit behavior of legitimate users as an inference (exploratory) attack and train a deep neural network to decide when to jam data transmissions. In addition, an adversary can jam transmissions during the sensing period to manipulate input data to training and testing processes of a machine learning classifier in terms of poisoning (causative) and evasion attacks, respectively  \cite{Yi18:Poisoning, Sagduyu19:Poisoning, Sagduyu:IoT}. Poisoning attack was also studied for cooperative spectrum sensing, where some of nodes may be malicious and provide wrong sensing results to the machine learning classifier deployed at a fusion center \cite{Zhuo2019}. On the other hand, the manipulation of test inputs to a machine learning-based modulation classifier was considered as an evasion attack in \cite{LarssonAML, Headley19, Deniz19, Silvija19, Kim2020, Kim2020-2, Kim2020-3}. The adversary can also launch a Trojan attack by manipulating training data to insert Trojans (triggers) and activating them later in test time \cite{Trojan}.

This paper presents a novel \emph{spoofing attack} built upon adversarial machine learning. This attack is based on training a deep neural network that generates synthetic wireless signals (namely, spoofing signals) that cannot be reliably distinguished from signals that are originated from intended users (e.g., legitimate users or higher priority users such as primary users). To generate such signals, the adversary uses a special kind of generative model, namely a \emph{generative adversarial network} (GAN) \cite{Goodfellow2014}, that is trained to learn to synthesize data samples that are statistically similar to real data samples. Our goal in this paper is to use the GAN from a wireless attack point of view and train it by an adversary pair, consisting of a transmitter and a receiver, collaborating over the air to spoof wireless signals such that the GAN-generated signals cannot be reliably discriminated from intended signals. Acting as the defender, there is a receiver that uses a deep neural network to classify signal sources as an intended transmitter or not, based on spectrum sensing results. The adversary generates spoofing signals to fool the classifier at the receiver into incorrectly authenticating its transmissions as intended. The preliminary version of the GAN-based spoofing attack was studied in \cite{Shi19:wiseml} for the special case of single antenna systems. In this paper, we consider the general case of multiple antenna capabilities at the defender and the adversary. In addition, we present the practical implementation of the GAN-based spoofing attack on embedded platforms and discuss the impact of network topology in more detail.

There are unique challenges for adversarial machine learning attacks in wireless communications. Unlike applications in other domains such as computer vision and NLP, data in wireless medium is received through channel effects (from the intended transmitter or the adversary to the receiver) and embedded with waveform and radio hardware effects that all need to be matched by the GAN.

As each device introduces its own phase shift and each channel has its own propagation gain and phase shift, a deep learning-based classifier is built at a receiver $R$ by collecting spectrum sensing results for signals from an intended transmitter $T$ and other signals. We show that this classifier is highly successful in distinguishing signals of $T$ from other random signals. To establish the baseline, we start with conventional spoofing attacks to fool this classifier at $R$. We can regard random signals from other transmitters as a naive spoofing attack. The success probability (the probability that signals from another transmitter is classified as from $T$) of this naive attack is only $7.8\%$ for the single antenna case. Even with multiple antennas, the success probability is still small (not more than $14.71\%$ that is achieved when four antennas are used by the transmitter). Then, we consider the replay attack, where an adversary transmitter $A_T$ amplifies and forwards the previously received signal from $T$, as a better spoofing attack, since it keeps some pattern of $T$ (but not the entire). We show that the attack success probability increases to $36.2\%$ when the replay attack is launched in the single antenna case. With multiple antennas used at the adversary transmitter, the attack success probability improves (e.g., to $69.8\%$ when all nodes use four antennas) but it is still much less than the success probability of the GAN-based spoofing attack that is introduced in this paper.

The adversary transmitter $A_T$ seeks to generate a signal that is statistically similar to the one received by the receiver $R$. However, it is challenging to generate such a signal without any knowledge on $T$'s waveform, phase shift, and the unique channel between $T$ and $R$. To overcome this challenge, we introduce a GAN-based approach to capture the cumulative effects from the observed signals, learn its distribution while training the GAN, and generate spoofing signals by the trained generator of the GAN. The adversary transmitter $A_T$ and its surrogate receiver $A_R$ (used only for training) train the GAN. In the training process, $A_T$ adds flags to its transmissions to inform $A_R$ of the true label. Using the received signals, $A_R$ trains a deep neural network as the discriminator of the GAN to classify signals as from $T$ or not, and sends the classification results back to $A_T$ as a feedback. Then, $A_T$ updates the generator of the GAN to generate better synthetic data, namely to increase the classification error probability at $A_R$. Thus, $A_T$ and $A_R$ iteratively play a minimax game, which trains a GAN to improve both the generator and the discriminator. Then, the generator at $A_T$ is used to generate high fidelity synthetic data samples (similar to real signals) by inherently capturing all waveform, channel, and device (radio hardware) effects jointly. The GAN-based spoofing attack increases the attack success probability  to $76.2\%$ for the single antenna case.

We consider multiple antennas for $A_T$ to introduce additional degrees of freedom when the spoofing signal is optimized. We show that the success probability of the GAN-based spoofing attack increases to $88.6\%$ and $100\%$ when two and four antennas are used, respectively, at $A_T$. Moreover, we consider different settings of multiple-input and multiple-output (MIMO) communications and show that more receiver antennas (at $R$) increase the classification accuracy at $R$ while more transmitter antennas (at $T$) can decrease this accuracy. On the other hand, the success probability of spoofing attacks increases with more transmitter antennas at either $T$ or $A_T$ while it decreases with more antennas at receivers ($R$ and $A_R$). We also show that this spoofing attack remains successful for different topologies and even when $A_T$ moves after training.

To demonstrate the practicality of the proposed spoofing attack, we present the implementation of the deep neural network structures developed for the receiver classifier and the spoofing signal generator of the GAN on two different embedded platforms, namely an embedded  graphics processing unit (GPU) and a field-programmable gate array (FPGA). For all antenna configurations considered, the latency is measured to be less than one millisecond (msec) on both platforms. The FPGA achieves lower latency for all antenna configurations compared to embedded GPU. In particular, the ratio of latency improvement by FPGA over embedded GPU is up to $36$ when the generator is run at $A_T$ equipped with four antennas.

The rest of the paper is organized as follows. Section~\ref{sec:related} discusses the related work. Section~\ref{sec:scenario} presents the system model. Section~\ref{sec:classifier} describes the pre-trained classifier to detect intended transmissions. Section~\ref{sec:adversary} presents the GAN-based spoofing attack, compares it with the replay attack, and then extends the setting to the MIMO case. Section~\ref{sec:implementation} presents the implementation of the GAN-based signal spoofing on embedded platforms. Section~\ref{sec:conclusion} concludes the paper.

\section{Related Work}
\label{sec:related}

There are different ways to attack wireless communications \cite{Clancy08:CogSec}. Spectrum sensing can be attacked in various forms including spectrum sensing data falsification (SSDF) \cite{Penna, Sagduyu2014}, primary user emulation (PUE) \cite{PUE}, eavesdropping \cite{Zou15:eavesdropping}, and noncooperation \cite{Sagduyu09:noncoop}. On the other hand, data transmissions can be jammed \cite{Sagduyu118:satellite, JamRes, UserCentric} via denial-of-service (DoS) attacks \cite{DoS} using different levels of prior information at the adversary \cite{Sagduyu11:jamming}. Separately, higher layer protocols can be also attacked, e.g., by manipulating routing at the network layer \cite{Lu2017} and inferring network flows \cite{LuCliff2017}.

Deep learning has been studied to secure wireless communications, such as authenticating signals \cite{Saad2018, Nousain18, Davaslioglu19}, detecting and classifying jammers of different types \cite{Poor2018, Wu2017, gunes},
and controlling communications to mitigate jamming effects \cite{Poor2018, UserCentric}. As cognitive radio capabilities are integrated into wireless communications, adversaries such as jammers become smarter, as well \cite{Poor2018, EnergyHarvestingCN}. In particular, deep learning was used to jam wireless communications building upon adversarial machine learning techniques \cite{Yi18:Jamming, Tugba2018, Yi18:Poisoning, Sagduyu19:Poisoning, Sagduyu:IoT}. Using wireless sensors, deep learning was also used to infer private information in analogy to exploratory attacks \cite{Liang18}.

Wireless attacks such as jamming and PUE attacks benefit from realistic spoofing signals such that the underlying attacks cannot be easily detected. Wireless spoofing attacks have been extensively studied \cite{Lichtman16:spoofing, Gai17:spoofing, Chen07:spoofing, Sheng08:spoofing, Sajjad18:spoofing}. Algorithms to spoof, jam, and sniff wireless signals were developed in \cite{Lichtman16:spoofing}. The optimal power distribution for a spoofing attack was derived in \cite{Gai17:spoofing}. As a countermeasure to spoofing attacks, \cite{Chen07:spoofing,Sheng08:spoofing} used feature-based methods based on the received signal strength (RSS), whereas \cite{Sajjad18:spoofing} trained a deep neural network.

In this paper, the adversary uses a special kind of generative model, namely a GAN \cite{Goodfellow2014}, to synthesize data samples that are statistically similar to real data samples. While GANs have been extensively used in computer vision and NLP applications, the use of GANs in wireless domain is in its early stage with emerging applications in spectrum sensing \cite{Kemal2018}, jamming/anti-jamming \cite{Tugba2018}, LTE signal generation \cite{Roy2019}, and IoT device fingerprinting \cite{Merchant2019}. In these studies, the GAN structure was often trained offline and centralized, i.e., there is no separate classifier at the target receiver, and the generator and the discriminator of the GAN are co-located at a single node position without accounting for relative positions of the transmitter-receiver pair and the adversary. This GAN setting is useful for training data augmentation by having full control and observation of transmitter and receiver signals, respectively. However, it does not provide the needed capability for the adversary to capture signal characteristics observed at the defender receiver.

We design the adversarial capability to spoof wireless signals with the GAN that captures not only waveform and channel effects but also device-related effects such as phase shift and relative positions of transmitters and receivers with respect to the adversary and the defender. For that purpose, the generator and the discriminator of the GAN need to be distributed to different locations. Those aspects were missing in the past applications of GANs to model wireless communication channels, e.g., \cite{Yang19:channel, Oshea18:channel, Oshea18:channel2}, where the GAN is centrally trained offline by accounting for waveform effects over a single channel only (without distinguishing the roles of  the transmitter and the receiver, and their relative channel effects).
The defender also uses a deep neural network but the main focus is on the use of deep neural networks at the adversary that jointly trains the discriminator and the generator of signal spoofing  for its own adversarial purposes.

Similar to wireless signal spoofing, adversarial perturbations can be added to wireless signals to fool signal classifiers such as the modulation classifier \cite{LarssonAML, Headley19, Deniz19, Silvija19}. The spoofing attack proposed in our paper has the same purpose as those evasion attacks, but it takes into account the channel effects, hardware effects (e.g., phase shifts) and relative positions of the adversary and defender, and does not require the adversary to synchronize the superposition of its signals with the adversary, which is hard to achieve in practice. In addition, adversarial training methods such as randomized smoothing \cite{certified} cannot be readily applied to mitigate the proposed spoofing attack, as they assume that perturbations are directly added to real signals.

We use the replay attack as a baseline for comparison purposes. The replay attack records the signal from an intended user transmission, and then amplifies and forwards it as a means of signal spoofing \cite{Kinnunen17:replay, Hoehn16:replay}. The replay attack is simple to launch as it maintains some features in the original signal, but it is not very effective since it cannot necessarily match the signal characteristics expected at the receiver and therefore it is easier to detect. In addition, the replay attack does not provide a mechanism to train (tune or optimize) itself against a signal classifier. We refer the interested reader to \cite{Kinnunen17:replay,Hoehn16:replay} for bounds on replaying signals as the spoofing attack and the countermeasure to detect these attacks. We will show that the GAN-based spoofing attack presented in our paper outperforms the replay attack since it jointly captures waveform, channel, and hardware effects as expected at the defender receiver.

\section{System Model}
\label{sec:scenario}
\begin{figure}
	\centering
	\includegraphics[width=\columnwidth]{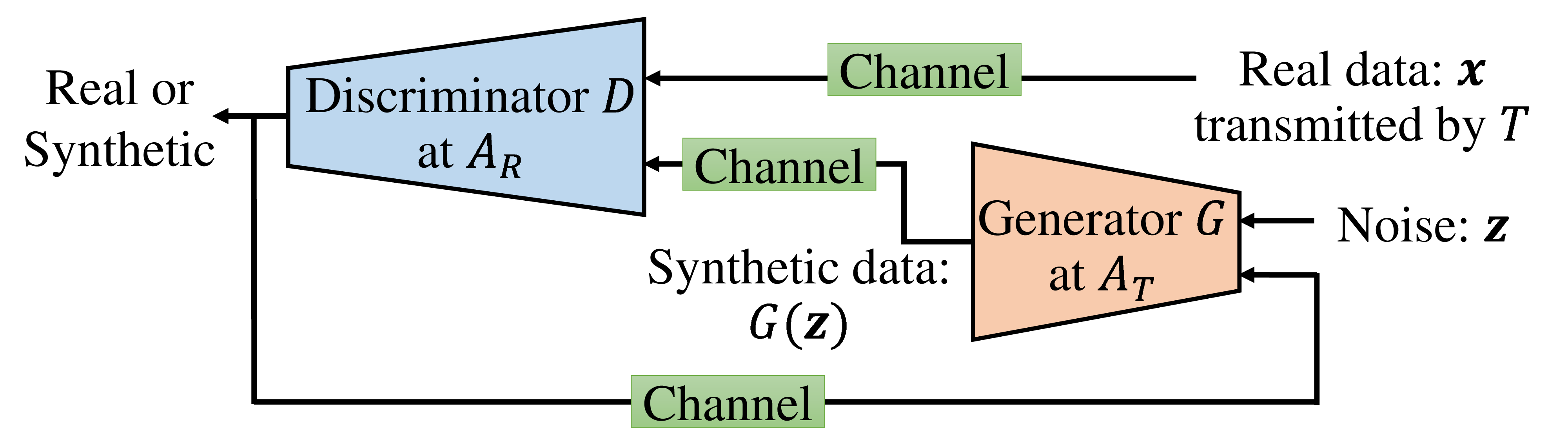}
	\caption{GAN structure trained for spoofing attack.}
	\label{fig:gan}
\end{figure}

There is one transmitter $T$ as a legitimate user and its corresponding receiver $R$. There is also an adversary pair of transmitter $A_T$ and receiver $A_R$.
The goal of the adversary is to generate spoofing signals that are misclassified by $R$ as signals transmitted from $T$. A pre-trained deep learning-based classifier is used at $R$ to determine whether a transmission is from $T$ or not. Since there are unique device properties (such as the phase shift) and communication channel properties (such as the channel gain) associated with signals from $T$, a random signal transmission by the adversary can be easily detected as an unintended transmission, as we show in Section \ref{sec:classifier}. Therefore, the goal of the adversary is to learn the unique pattern embedded in $T$'s signals received at $R$ and generate spoofing signals  following the same pattern. $A_T$ and $A_R$ jointly train a GAN (see Fig.~\ref{fig:gan}), namely $A_T$ trains the generator of the GAN and $A_R$ trains the discriminator of the GAN. Note that all transmissions go through wireless channels. Once the GAN is trained, only the generator is used to generate spoofing signals in test (inference) time of the spoofing attack.
We do not assume that the adversary has any knowledge of $T$'s coding/modulation scheme or the channel between $T$ and $R$.
Instead, the adversary needs to learn their combined effect online through the collaboration of $A_T$ and $A_R$.

\begin{figure}
	\centering
	\includegraphics[width=0.75\columnwidth]{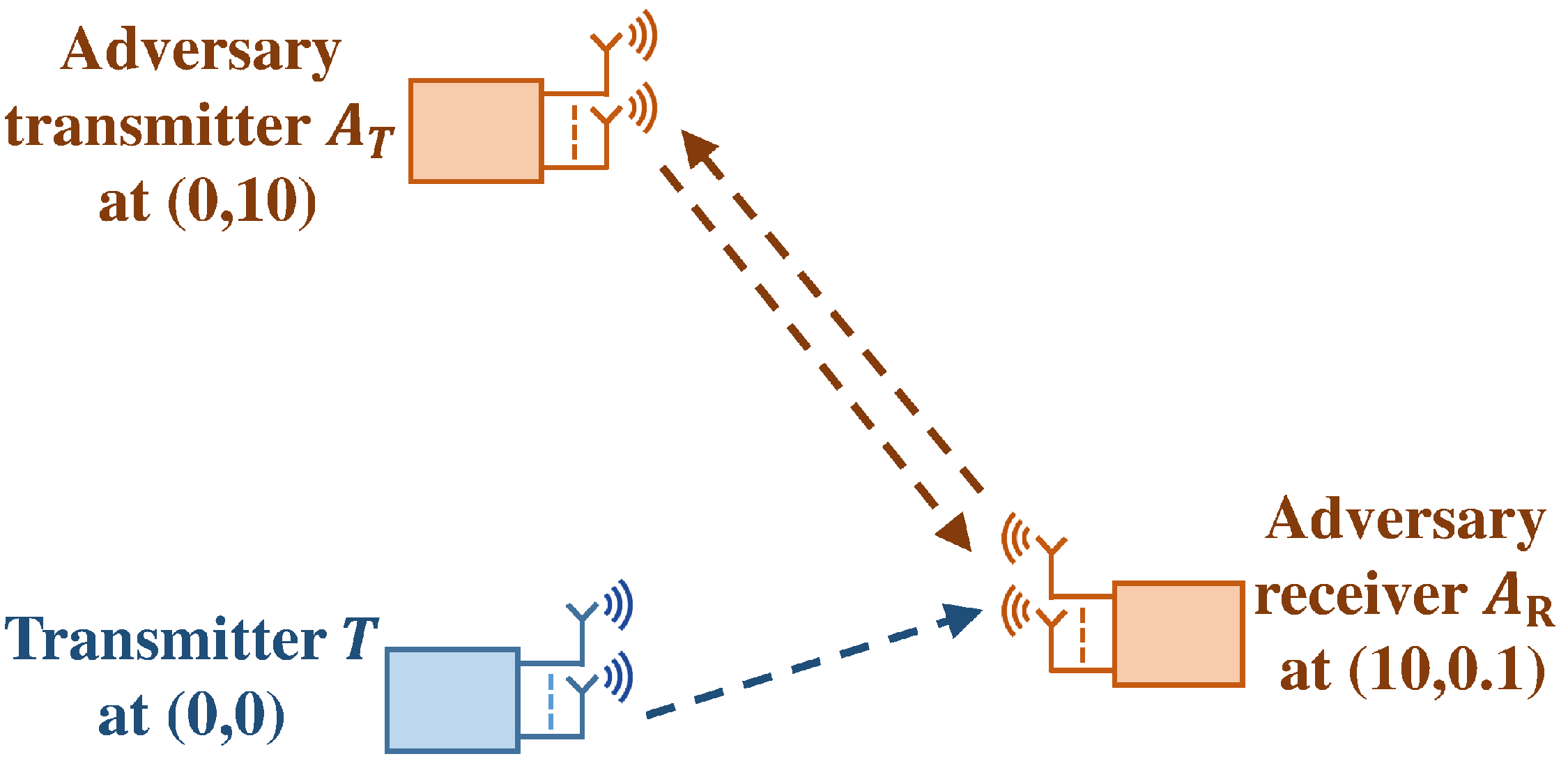}
	\caption{Network topology during the training process for spoofing attack.}
	\label{fig:topology1}
\end{figure}

Each node may have potentially multiple antennas. Suppose that $T$ has $N_T$ antennas, $R$ has $N_R$ antennas, $A_T$ has $N_A$ antennas, and $A_R$ has $N_R$ antennas. Note that $A_R$ needs to collect similar data as $R$, and thus we assume that $A_R$ has the same number of antennas as $R$.
We assume additive white Gaussian noise and normalize all powers with respect to noise power. The transmit power of $T$ is $P=1000$.
We assume that there is a device-related phase shift for transmissions of $T$ and this device property is unknown to $A_T$ and $A_R$. The channel between any two nodes is modeled by Rayleigh distribution.

We consider three types of spoofing attacks.
\begin{itemize}
\item [1.] \emph{Random signal attack:} $A_T$ transmits random signals with power $P$.

\item [2.] \emph{Replay attack:} $A_T$ records signals from $T$, and then amplifies and forwards it to $R$. Since $A_T$ does not have any knowledge on channel gain, it cannot optimally tune its power. Thus, we assume $A_T$ uses fixed power $P$ to amplify signals.

\item [3.] \emph{GAN-based spoofing attack:} $A_T$ uses a GAN to generate synthetic signals and transmits them with power up to $P$.
\end{itemize}

When $A_R$ is placed close to $R$ such that the channel from $T$ (or $A_T$) to $A_R$ is similar to the channel from $T$ (or $A_T$) to $R$, signals received by $A_R$ are similar to signals received by $R$. Hence, if $A_R$ cannot distinguish whether a signal is from $T$ or $A_T$, $R$ cannot either. In Section~\ref{sec:GAN-based}, we will show that the GAN-based spoofing attack is more successful than other spoofing attacks even when $A_R$ is located far away from $R$.

When $A_T$ transmits a spoofed signal, its adds a flag (i.e., the true label) to inform $A_R$ that this signal is from $A_T$. Using real and synthetic signals received along with labels through flags from $A_T$,
$A_R$ starts training a discriminator to classify signals from $T$ or $A_T$, and informs $A_T$ of its classification results as a feedback (see Fig.~\ref{fig:gan}). We assume that this $1$-bit feedback is transmitted with a strong channel code to ensure that the correct feedback can be decoded at the generator side.

\begin{figure}
	\centering
	\includegraphics[width=0.75\columnwidth]{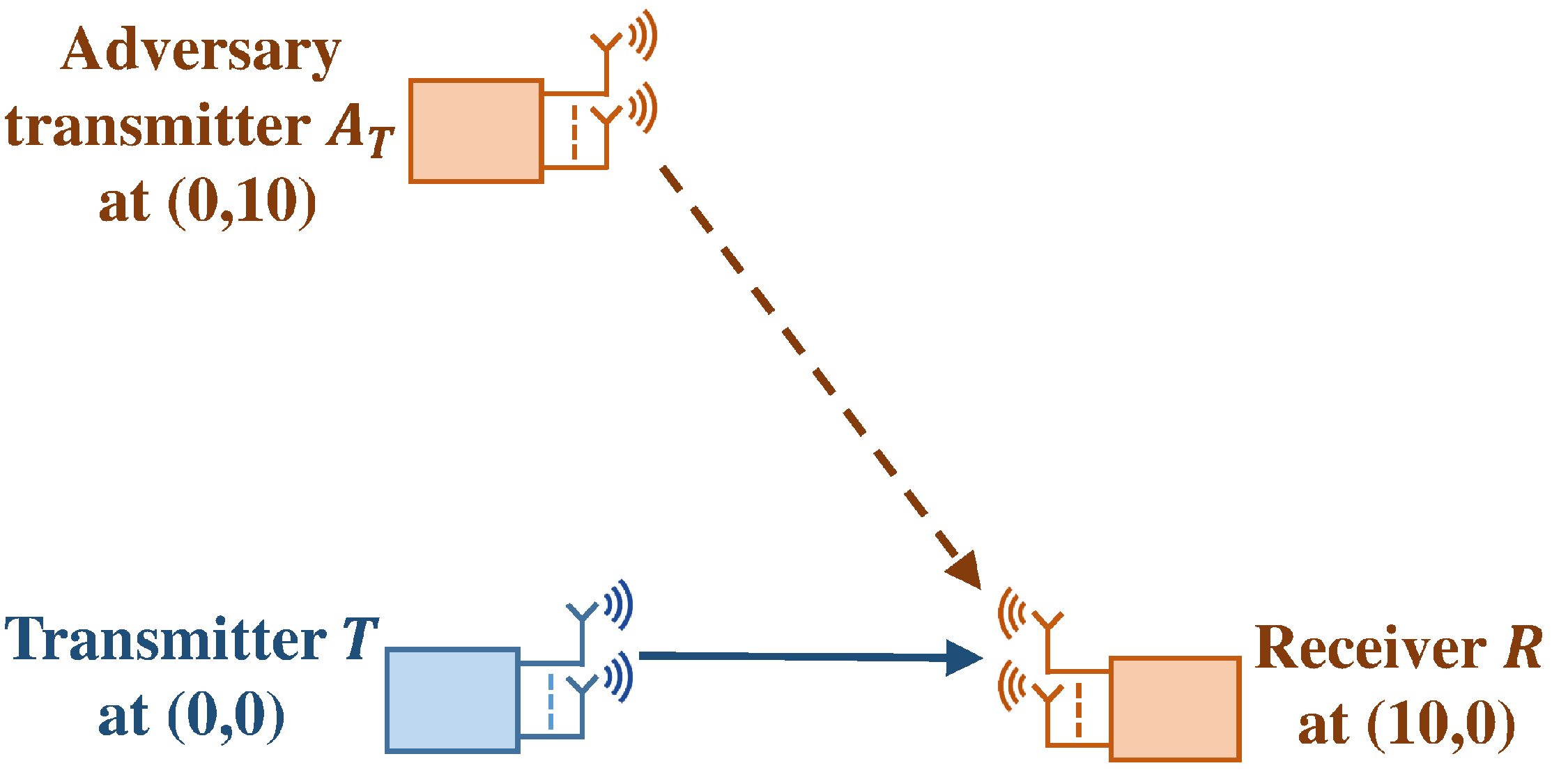}
	\caption{Network topology during the spoofing attack.}
	\label{fig:topology2}
\end{figure}

In the meantime, $A_T$ starts training a generator to enhance its synthetic signals and make them  statistically more similar to $T$'s signals such that the classification error at $A_R$ should increase. $A_T$ and $A_R$ continue with this process until convergence. In this setting, $A_T$ and $A_R$ play a minimax game, which corresponds to the GAN process (see Fig.~\ref{fig:topology1}), but it is played over the air. When the GAN converges, the generator at $A_T$ should be able to generate synthetic signals that are statistically very similar to signals of $T$ received by $R$, and then it is used for the spoofing attack (see Fig.~\ref{fig:topology2}).

In the next two sections, we will describe in detail the classifier at receiver $R$ and the GAN at adversary transmitter $A_T$ and receiver $A_R$.

\section{The Classifier for Signal Authentication}
\label{sec:classifier}

Receiver $R$ has a pre-trained classifier (a deep neural network) to distinguish whether a received signal is from $T$, or not. The classifier takes spectrum sensing results as input. $R$ senses the channel for a short period of time that corresponds to $8$ bits of data transmission. We assume that $T$ uses QPSK modulation for its transmissions, although the classifier and the GAN in the next section do not assume any knowledge of modulation scheme (or other waveform characteristics) used by the target transmitter. Under QPSK, there are four possible modulated signals, where each signal may have a different phase shift for $2$ bits. Each antenna adds its own phase shift, as well. Denote $\theta_{Ti}$ as the phase shift of the $i$-th antenna at $T$, which is added to the QPSK signal's phase shift. Note that other settings on the number of bits in sensing data and modulation type  can also be used without changing the algorithms (classifiers) developed in this paper. In this setup, each input data sample of $R$ consists of four received signals. Each signal is uniformly sampled $100$ times. Therefore, each input data sample of $R$ consists of $400$ I/Q features when we consider the single antenna at $R$. On the other hand, when multiple antennas are used at $R$, the number of data features increases, e.g., there are $1600$ I/Q features for each sample when $R$ has $4$ antennas.

As an example of $T$'s signals, QPSK determines a phase shift $\frac{\pi}{4}$ for the coded signal when $T$ transmits two bits $0$ and $0$.
Adding $\theta_{Ti}$ and a random channel phase shift $\theta_{Ti, Rj}$ (from the $i$-th antenna at $T$ to $j$-th antenna at $R$) under the Rayleigh model, the received signal has phase shift $\frac{\pi}{4} + \theta_{Ti} + \theta_{Ti, Rj}$. The $k$-th sample point, $0 \le k < 100$, has phase shift $\frac{\pi}{4} + \theta_{Ti} + \theta_{Ti, Rj} + \frac{k \pi}{50}$.
The received power is $g_{TR} \frac{P}{N_T}$, where $g_{TR}$ is a random channel gain from $T$ to $R$ under the Rayleigh model and $N_T$ is the number of antennas at $T$. Note that we approximately assume that the channel gain is independent of the antennas at $T$ and $R$. We assume that the mean value of channel gain is $d^{-2}$, where $d$ is the distance between a transmitter and a receiver, although the classifier and the GAN in the next section do not assume any knowledge of channel gain model.
In this setting, the $k$-th sampled data at $R$'s $j$-th antenna is given by
\begin{eqnarray}
d_{T,Rj}^k = g_{TR} \frac{P}{N_T} \sum_{i=1}^{N_T} e^{j (\frac{\pi}{4} + \theta_{Ti} + \theta_{Ti,Rj} + \frac{k \pi}{50})} \; .
\end{eqnarray}
In the training process, a flag is sent by $T$ to indicate its transmissions and it is used to label samples. This way, $R$ collects a number of samples with labels to build the training data and trains its classifier that is used to predict signal labels (`$T$' or `not $T$'). This classifier involves two types of errors, namely \emph{misdetection} (the signal of $T$ is classified as from other transmitters) and \emph{false alarm} (the signal from other transmitters is classified as from $T$). Denote $e_{MD}$ and $e_{FA}$ as the probabilities of misdetection and false alarm at $R$, respectively. Then, the objective of $R$ is to minimize $\max\{ e_{MD}, e_{FA} \}$. Suppose that test data has $n$ samples, there are $N_T$ samples with signals from $T$, and in these samples there are $n_{MD}$ misdetections and $n_{FA}$ false alarms. Then, we have $e_{MD} = \frac{n_{MD}}{N_T}$  and $e_{FA} = \frac{n_{FA}}{n-N_T}$.

\begin{figure}
	\centering
	\includegraphics[width=\columnwidth]{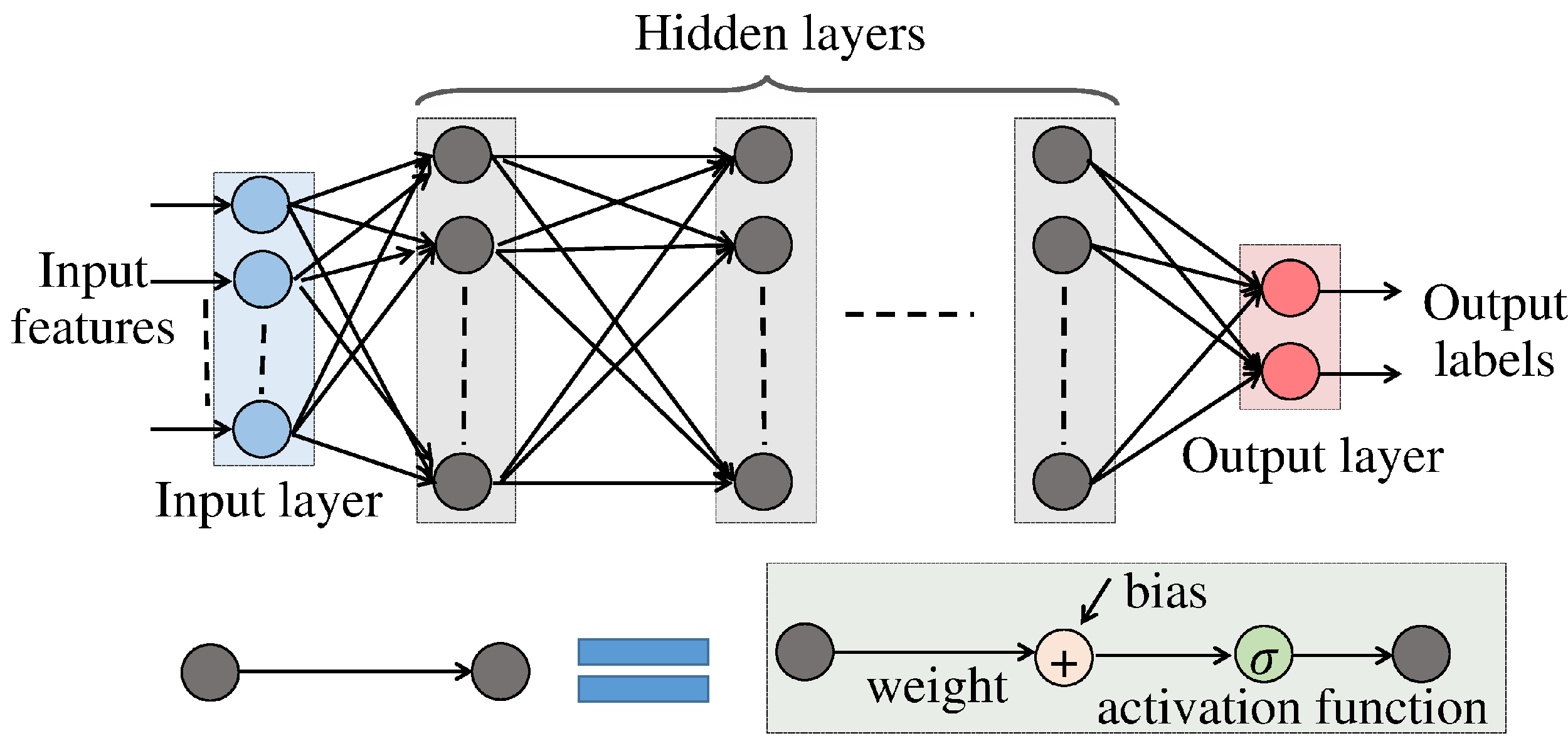}
	\caption{The structure of a feedforward neural network.}
	\label{fig:fnn}
\end{figure}

\begin{table*}
\caption{Classification errors (Misdetection, False alarm) under different MIMO settings.}
\label{table:classifier}
\small
\centering
\begin{tabular}{c|c|c|c|c} \hline
$N_T$ \textbackslash $N_R$ & $1$ & $2$ & $3$ & $4$ \\ \hline \hline
$1$ & $(7.34\%, 7.86\%)$ & $(6.24\%, 7.53\%)$ & $(5.91\%, 6.33\%)$ & $(5.89\%, 6.31\%)$ \\ \hline
$2$ & $(7.94\%, 12.90\%)$ & $(6.75\%, 10.97\%)$ & $(6.40\%, 10.43\%)$ & $(6.37\%, 10.38\%)$ \\ \hline
$3$ & $(8.79\%, 14.29\%)$ & $(7.47\%, 12.15\%)$ & $(7.09\%, 11.50\%)$ & $(7.06\%, 11.50\%)$ \\ \hline
$4$ & $(11.9\%, 14.71\%)$ & $(10.12\%, 12.95\%)$ & $(9.62\%, 12.46\%)$ & $(9.57\%, 12.42\%)$ \\ \hline
\end{tabular}
\end{table*}

In this paper, we use TensorFlow to train feedforward neural networks (see Fig.~\ref{fig:fnn}) as deep learning classifiers using cross-entropy as the loss function. For each antenna configuration, hyperparameters of the  deep neural network are selected to minimize $\max\{ e_{MD}, e_{FA} \}$ to balance the accuracy on each label. The default deep neural network structure of the classifier at $R$ is given as follows.
\begin{itemize}
	\item Number of neurons at the input layer depends on the antenna configuration (e.g., $400$ for $R$ and $1600$ for $4$ antennas at $R$).
	\item There are $3$ hidden layers.
	\item Each hidden layer has $50$ neurons.
	\item The output layer has $2$ neurons.
	\item The activation function at hidden layers is Rectified linear unit (ReLU).
	\item The activation function at output layer is softmax.
	\item Batch size is $100$.
	\item Number of training steps is $1000$.
\end{itemize}
With hyperparameter tuning, the above default values may change. For example, batch size is optimized to $150$ for the single-input and single-output (SISO) antenna case ($N_T =N_R = 1$).

For simulations, $T$ is located at $(0,0)$, $R$ is located at $(10,0)$, $A_T$ is located at $(0,10)$, and $A_R$ is located at $(10,0.1)$ (see Figs.~\ref{fig:topology1} and \ref{fig:topology2}). As training data, $R$ collects $1000$ samples, each with $400$ spectrum sensing results and label (`$T$' or `not $T$'), and runs the classifier on another set of $1000$ samples to evaluate the classifier accuracy. There are $504$ signals from $T$ and $496$ signals from other transmitters in the test data.

When we consider the SISO case, $39$ signals from other transmitters are identified as signals from $T$ and $37$ signals from $T$ are identified as other signals.
As a result, $e_{FA} = 39/496 = 7.86\%$, $e_{MD} = 37/504 = 7.34\%$ indicating that $R$ can distinguish signals of $T$ with small error. Note that this case can be interpreted as a naive spoofing attack (the adversary transmits random signals), where the attack success probability is only $7.86\%$.

For the general MIMO case, we vary $N_T$ (the number of antennas at $T$) and $N_R$ (the number of antennas at $R$). Results are shown in Table~\ref{table:classifier} where each data entry lists ($e_{MD}$, $e_{FA}$).
In general, larger $N_T$ means that $T$ can generate more complex transmit signals and thus the complexity of received signals increases. On the other hand, larger $N_R$ means that $R$ can collect more copies of received signals and thus more information can be collected. Therefore, error probabilities of the target classifier at the receiver increase with larger $N_T$ while error probabilities decrease with larger $N_R$. These observed trends indicate how degrees of freedom offered by multiple antennas drive signal classification and provide the basis for the results presented in the next section.

\section{Spoofing Attacks}
\label{sec:adversary}
The goal of adversary transmitter $A_T$ is to transmit signals that are similar to transmitter $T$'s signals and fool target receiver $R$ into classifying signals from $A_T$ as signals from $T$. In the previous section, we considered the naive spoofing attack, where $A_T$ transmits random signals. In this section, we first present the replay attack, where $A_T$ records $T$'s transmissions and replays them later. Then, we introduce the GAN-based spoofing attack, where the adversary consists of two nodes, transmitter $A_T$ and receiver $A_R$. We assume the worst case from the adversary point of view that $A_T$ and $A_R$ do not know $T$'s waveform or device-related phase shifts, or channels between $T$ and $R$.

\subsection{Replay Attack based on Amplifying and Forwarding Signals}

\begin{table*}
\caption{Success probability (\%) of replay attack under different MIMO settings.}
\label{table:succ-replay}
\small
\centering
\begin{tabular}{c|c|c|c|c} \hline
$N_T$ \textbackslash $N_R$ & $1$ & $2$ & $3$ & $4$ \\ \hline \hline
$1$ & $36.2,47.6,42.8,45.8$ & $35.6, 38.8, 42.2, 43.4$ & $34.2, 33, 40.2, 40.2$ & $33.2, 32.2, 32.4, 37.4$ \\ \hline
$2$ & $52.6, 56.6, 58, 59.4$ & $51.8, 55, 56.4, 58$ & $43.2, 51.2, 51, 52$ & $42.8, 45.6, 48.4, 50.2$ \\ \hline
$3$ & $53.6, 57.6, 57.8, 57.8$ & $51.8, 55.4, 56.2, 58.2$ & $45, 55.8, 56, 58$ & $44.8, 45.4, 48, 50$ \\ \hline
$4$ & $69.4, 70.8, 69.8, 71.2$ & $66.8, 68, 69, 69.6$ & $62.8, 63.4, 69.4, 70.2$ & $61.4, 62.8, 67, 69.8$ \\ \hline
\end{tabular}
\end{table*}

The replay attack is based on simply amplifying and forwarding signals, i.e., $A_T$ receives signals from $T$, records them, amplifies to power $P$, and forwards them to $R$. Denote $\theta_{A_{Ti}}$ as the phase shift for the $i$-th antenna at $A_T$, $\theta_{A_{Ti},A_{Rj}}$ as the phase shift from the $i$-th antenna at $A_T$ to the $j$-th antenna at $A_R$, and $g_{ij}$ as the random channel gain for the Rayleigh channel from node $i$ to node $j$. We assume that none of $T$'s parameters are known to $A_T$.
As an example of replay attack, consider two bits $0$ and $0$ transmitted by $T$ that determines the phase shift of $\frac{\pi}{4}$ for QPSK.
The signals received by $R$ has the phase shift of $\frac{\pi}{4} + \theta_{Ti,A_{Th}} + \theta_{A_{Th}} + \theta_{A_{Th} Rj}$ for the signal transmitted by the $i$-th antenna at $T$, forwarded by the $h$-th antenna at $A_T$, and received by the $j$-th antenna at $R$.
The received power is $g_{A_T R} \frac{P}{N_A}$, when $A_T$ has $N_A$ antennas.
Then, the $k$-th sampled data at $R$'s $j$-th antenna is given by
\begin{eqnarray}
&& \hspace{-8mm} d_{A_T, Rj}^k \nonumber \\
&\hspace{-8mm} =& \hspace{-6mm} g_{A_T,R} \frac{P}{N_A} \hspace{-1mm} \sum_{i=1}^{N_T} \sum_{h=1}^{N_A} \hspace{-0.5mm} e^{j (\frac{\pi}{4} + \theta_{Ti} + \theta_{Ti, A_{Th}} \hspace{-0.5mm} + \theta_{A_{Th}} \hspace{-0.5mm} + \theta_{A_{Th}, Rj} + \frac{k \pi}{50})}.
\label{eq:dar}
\end{eqnarray}
If the same signal is transmitted by $T$, the $k$-th sampled data at $R$'s $j$-th antenna is given by
\begin{eqnarray}
d_{T,Rj}^k = g_{TR} \frac{P}{N_T} \sum_{i=1}^{N_T} e^{j (\frac{\pi}{4} + \theta_{Ti} + \theta_{Ti,Rj} + \frac{k \pi}{50})} \; \label{eq:dtr}.
\end{eqnarray}
While both received power and phase shift in (\ref{eq:dar}) and (\ref{eq:dtr}) are different, a simple detector based on discriminating received power and/or phase shift cannot be applied. Classification needs to be done with a limited number of data samples to detect the intruder at the physical layer.
However, the received power and the phase shift cannot be accurately estimated due to random channel. Therefore, either misdetection or false alarm probability is large if we set a small or large region, respectively, around the actual value of $T$.

When we simulate the replay attack in the setting described in the previous section, the success probability of spoofing attack is increased from $7.86\%$ (when $A_T$ transmits random signals) to $36.2\%$ for the SISO case (i.e., $N_T = N_R = N_A =1$). While some signal pattern from $T$ at $R$ is captured by amplifying and forwarding the recorded signals, signals from $T$ and signals forwarded by $A_T$ differ even for the same data and therefore $R$ can still successfully classify most of signals that are spoofed by the replay attack.

For the MIMO case, we can vary $N_T$ (the number of antennas at $T$), $N_R$ (the number of antennas at $R$), and $N_A$ (the number of antennas at $A_T$). Note that since $A_R$ needs to learn the received signal patterns at $R$, the number of antennas at $A_R$ should be the same as the number of antennas at $R$, i.e., $N_R$. The transmit power at $A_T$ is equally divided among antennas, since $A_T$ does not know $T$'s power allocation policy. We obtain results in Table~\ref{table:succ-replay}, where each data entry lists a sequence of numbers when $N_A$ varies from $1$ to $4$. Note that the first entry for $N_T = N_R = 1$ corresponds to the SISO case for the transmitter-receiver pair. We observed the following trends for the MIMO case. The attack success probability increases with larger $N_T$, since the transmit signal is more complex and thus is easy to attack. The attack success probability also increases with larger $N_A$, since the adversary can generate complex spoofing signals.
On the other hand, the attack success probability decreases with larger $N_R$, since the receiver can collect more data for its classifier.

\subsection{GAN-based Spoofing Attack} \label{sec:GAN-based}

The next attack relies on the collaboration of the adversary transmitter-receiver pair, $A_T$ and $A_R$, to run the GAN over wireless channels. Traditionally, a GAN is implemented by the same processor controlling both the generator and discriminator. Since $A_T$ needs the generator of the GAN to generate synthetic signals, the GAN can be implemented entirely at $A_T$.
However, this requires $A_R$ to collect features of its received signal and transmit features to $A_T$, which causes high overhead, creates a communication bottleneck, and makes the adversary easy to detect because of larger communication footprint. For a more realistic scenario, we propose to run the discriminator of the GAN separately at $A_R$ to avoid such overhead such that only limited feedback is needed ($A_R$ needs to send only one bit for the classification decision back to $A_T$).

We now present design details. $A_R$ collects signal samples from $T$ and  $A_T$, where $A_T$ flags its transmissions to inform $A_R$ of ground truth. Then, $A_R$ trains the first version of discriminator $D$ based on these data samples with the objective of minimizing the classification error, i.e.,
\begin{eqnarray}
\min_{D} \mathbb{E}_{\bm{z} \sim p_{\bm{z}}} [\log(1 - D(G(\bm{z})))] - \mathbb{E}_{\bm{x} \sim p_{data}} [\log(D(\bm{x}))] \; ,
\end{eqnarray}
where $\bm{z}$ is a noise input to generator $G$ with a random distribution of $p_{\bm{z}}$, $G(\bm{z})$ is the generator output and input data $\bm{x}$ has distribution $p_{data}$.

In the meantime, $A_T$ collects classification results from $A_R$, trains the first version of generator $G$ to generate synthetic data, and then transmits spoofing signals to $A_R$. The objective of $A_T$ is to maximize $A_R$'s classification error, i.e.,
\begin{eqnarray}
\max_{G} \mathbb{E}_{\bm{z} \sim p_{\bm{z}}} [\log(1 - D(G(\bm{z})))] - \mathbb{E}_{\bm{x} \sim p_{data}} [\log(D(\bm{x}))] \; ,
\label{eq:obj1}
\end{eqnarray}
where $D$ is the first version of discriminator. This process continues with updated versions of $G$ and $D$ trained over time in subsequent rounds. This way, $G$ and $D$ improve in each round until they converge. The entire process corresponds to a minimax game played between $A_T$ and $A_R$ as follows.
\begin{eqnarray}
\max_{G} \min_{D} && \mathbb{E}_{\bm{z} \sim p_{\bm{z}}} [\log(1 - D(G(\bm{z})))] \nonumber \\
&& - \mathbb{E}_{\bm{x} \sim p_{data}} [\log(D(\bm{x}))] \; .
\end{eqnarray}
Although traditionally a GAN is run at one entity, we split here $G$ to $A_T$ and $D$ to $A_R$, each under different channel effects. When $G$ is trained with the objective in (\ref{eq:obj1}), the gradients of $G$ rapidly vanish, which makes the training of GAN very difficult. To address the vanishing gradient problem, the objective function at $G$ is changed to the following \cite{Goodfellow2014}:
\begin{eqnarray}
\max_{G} \mathbb{E}_{\bm{z} \sim p_{\bm{z}}} [\log(1 - D(G(\bm{z})))].
\end{eqnarray}
Once the solution of $G$ and $D$ converges, $A_T$ runs $G$ to generate synthetic signals and transmits them to $R$. Then, signals of $A_T$ received by $R$ are statistically similar to the received signals from $T$. In summary, the GAN-based spoofing attack has the following steps.
\begin{enumerate}
\item Training process includes two interactive steps.
\begin{enumerate}
    \item $A_R$ collects data from $T$ and $A_T$. $A_R$ builds the discriminator $D$. $A_R$ sends classification results to $A_T$.
    \item $A_T$ receives classification results from $A_R$. $A_T$ builds the generator $G$. $A_T$ generates more data for $A_R$.
\end{enumerate}

\item Once training is complete, $A_T$ uses its generator $G$ to generate spoofing signals and transmits them.
\end{enumerate}

\begin{table*}
\caption{Success probability (\%) of GAN-based spoofing attack under different MIMO settings.}
\label{table:succ}
\small
\centering
\begin{tabular}{c|c|c|c|c} \hline
$N_T$ \textbackslash $N_R$ & $1$ & $2$ & $3$ & $4$ \\ \hline \hline
$1$ & $76.2,88.6,100,100$ & $64.8,75.4,84.8,85$ & $63.2,73.2,82.6,82.6$ & $62.0,72.0,81,81.2$ \\ \hline
$2$ & $90.2,95.6,100,100$ & $76.6,81.2,84.6,85$ & $74.6,79,82.2,82.4$ & $73.4,77.8,81.2,81.2$ \\ \hline
$3$ & $89.6,99.6,99.6,100$ & $76.2,84.4,84.6,84.8$ & $74.2,82,82.2,82.2$ & $73.8,80.8,80.8,81$ \\ \hline
$4$ & $95.4,99.6,99.4,99.4$ & $81,84.6,84.2,84.4$ & $78.8,82,82,82$ & $77.6,81,80.6,80.8$ \\ \hline
\end{tabular}
\end{table*}

For simulations, we assume that as the first step, $A_R$ collects $500$ signal samples from $T$ and $500$ signal samples from $A_T$. Each sample has coded data of $8$ bits under the QPSK modulation, i.e., $4$ coded signals.
The sampling rate for a signal is $100$, and thus the total data (features) for a sample is $400$.

Both $G$ and $D$ have three hidden dense layers, each with $128$ neurons. The input layer of $G$ has 100 neurons. The output layer of $G$ has $400$ neurons per antenna at $R$ (e.g., $400$ for $N_R = 1$ and 1600 for $N_R = 4$), which is also the size of the input layer of $D$. The output layer of $D$ has $2$ neurons. The rest of the hyperparameters is the same as the deep neural network used for the classifier at $R$ (see Section \ref{sec:classifier}).

\begin{table}
\caption{The impact of $A_T$'s location on success probability of GAN-based spoofing attack.}
\label{table:at}
\small
\centering
\begin{tabular}{c|c} \hline
$A_T$ location & Success probability \\ \hline \hline
$(0,5)$ & $98.6\%$ \\ \hline
$(0,10)$ & $76.2\%$ \\ \hline
$(0,15)$ & $75.6\%$ \\ \hline
$(0,20)$ & $54.6\%$ \\ \hline
\end{tabular}
\end{table}

We follow the simulation setting of Section~\ref{sec:classifier}. We assume that the GAN converges when the maximum perturbation in $G$ and $D$ loss functions over the most recent $100$ epochs of the GAN training drops below $5\%$ of current loss value. With this convergence criterion, we measure that the GAN is run only for $478$ epochs. The attack success probability of the GAN-based spoofing attack is $76.2\%$ for the SISO case ($N_T = N_R = N
_A = 1$). In other words, the classifier at $R$ (which works very well to discriminate signals of $T$ from random or replayed signals) cannot successfully discriminate synthetic signals generated by the GAN. The complete results of spoofing attack under different MIMO settings are shown in Table~\ref{table:succ}, where each data entry lists a sequence of numbers when $N_A$ varies from $1$ to $4$. We observed the same trends with the use of multiple antennas as in the case of replay attacks, namely the attack success probability increases with larger $N_T$ or $N_A$, and decreases with larger $N_R$. Comparing Tables~\ref{table:succ-replay} and \ref{table:succ}, we can see that the GAN-based spoofing attacks are much more effective than replay attacks.

The above results are obtained when $A_T$ is located at $(0,10)$. We now study the impact of $A_T$'s locations. We set $A_T$'s location as $(0, 5)$, $(0, 15)$ and $(0, 20)$, respectively, and keep all other settings unchanged. We focus on the SISO setting in this study. The success probability of the GAN-based spoofing attack is shown in Table~\ref{table:at}. We can see that if $A_T$ is close to $T$, i.e., $A_T$ is at $(0, 5)$, it can generate high-fidelity synthetic signals for the spoofing attack. As a result, the attack success probability is very high ($98.6\%$). On the other hand, if $A_T$ is far away from $T$, e.g., $A_T$ is at $(0, 20)$, there is a significant difference between its channel to $R$ and $T$'s channel to $R$. Moreover, $A_T$'s power is limited, and thus $A_T$ may not be able to compensate its channel propagation gain. Now, the attack success probability is reduced ($54.6\%$), but it is still much higher than the success probability by using random signals ($7.86\%$). If $A_R$ is located very close to $R$, it can observe similar signals as $R$ and the spoofing attack is very likely to be successful. If $A_R$ is not close to $R$, the attack success probability may drop. To show this, we move $A_R$ from $(10, 0.1)$ to $(10, 10)$ and check multiple locations while keeping all other settings unchanged. The success probability of the GAN-based spoofing attack is between $60.6\%$ and $97.8\%$, which is always much better than spoofing with random signals or replay attack. Although the location of $A_R$ plays a significant role regarding the success probability, this impact is complex. The success probability of spoofing attack is not monotonically decreasing when the distance between $R$ and $A_R$ increases.

Finally, we study how the attack success changes if the network topology changes from training time to test time. For that purpose, we move $A_T$ from $(0,10)$ to a new position after the training process. $A_T$ can still use its current generator to launch the attack. Table~\ref{table:mobility} shows results under different $A_T$ locations, where we focus on the SISO setting and assume that the GAN is not retrained. We observe that as $A_T$ moves away from $T$, the distribution of the received signal changes and the attack success probability decreases. However, the attack success probability is still significantly higher than the one achieved by the replay attack when $A_T$ does not move.
If $A_T$ moves far away from its position when training, it is expected that the attack cannot be very successful. In that case, $A_T$ may request $A_R$ to retrain the GAN together, and then run the retrained generator for spoofing attack. Other topology changes, such as moving $T$ or $R$, will provide similar results. Thus, the GAN-based spoofing attack can be applied in mobile scenarios.

\begin{table}
\caption{The impact of $A_T$'s mobility (after training) on success probability of GAN-based spoofing attack.}
\label{table:mobility}
\small
\centering
\begin{tabular}{c|c} \hline
$A_T$'s location & Success probability \\ \hline \hline
$(0,10)$ & $76.2\%$ \\ \hline
$(0,11)$ & $65.2\%$ \\ \hline
$(0,15)$ & $61.0\%$ \\ \hline
$(0,20)$ & $56.2\%$ \\ \hline
\end{tabular}
\end{table}

\section{Practical Implementation}
\label{sec:implementation}
In this section, we evaluate the processing time associated with signal spoofing and classification operations. For that purpose, we have implemented each deep learning task (classifier at the receiver, generator at the adversary's transmitter, and discriminator at the adversary's receiver) on two embedded platforms, NVIDIA Jetson Nano Developer Kit \cite{nano} and Xilinx Zynq UltraScale+ XCZU9EG FPGA \cite{xilinx}.

The trained software model of the GAN in Keras is converted to a TensorFlow graph to create an inference graph with TensorRT \cite{tensorRT}, inference optimizer for NVIDIA's embedded GPU systems. For an efficient deployment, the model is quantized by TensorRT to a 16-bit fixed point (FP16) implementation that effectively reduces the memory consumption. The tensors at each layer are fused together to optimize the use of embedded GPU memory and bandwidth.

For the FPGA implementation, Vivado Design Suite \cite{Vivado} is used to simulate, and then synthesize the FPGA code. To export the trained Keras model to Vivado, the model is again quantized to FP16. The weights and bias of each layer are converted to the FPGA-readable format following the approach in \cite{FPGAmilcom} and interfaced with Vivado. For the timing analysis, the primary clock operates with 100~MHz cycle. Static timing analysis validates that there is enough timing margin (slack) to make setup and hold calculations. This way, it is ensured that there is sufficient margin in placement and routing such that the data arrives and remains valid (stored in the register and ready to be used) before the clock transitions and held valid for a period of time after the transition.

The inference tests are repeated 1000 times and the average inference time (latency to process one sample) is computed. The latency results are shown in Tables~\ref{table:gpu} and \ref{table:fpga} for embedded GPU and FPGA, respectively. Since we assume that the classifier at $R$ and the discriminator at $A_R$ have the same deep neural network structures, their inference times are the same. On the other hand, the deep neural network structure of the generator at $A_T$ is larger than others so the corresponding latency is higher. In all cases, latency is lower when a single antenna is used at each node, since the underlying deep neural network is smaller compared to the MIMO extension (multiple antennas are used at $R$ or $A_T$). Note that the number of features or the size of first (input) layer at $R$ increases linearly with the number of antennas at $R$ while the size of generated samples by $A_T$ or the size of last (output) layer at $A_T$ increases linearly with the number of antennas at $A_T$. Moreover, a larger input/output layer also yields larger deep neural networks.
Overall, FPGA achieves order-wise (up to $36$ times) smaller latency (measured at the microsecond level) compared to embedded GPU as it operates with faster cycle, while latency achieved by embedded GPU is still less than a millisecond (as a time reference, note that a typical frame in the IEEE 802.11ac standard is 5.484 msec, which is much larger than the processing times measured).

\begin{table}
\caption{Inference time of deep neural networks of the receiver and the adversary on Embedded GPU.}
\label{table:gpu}
\small
\centering
\begin{tabular}{c|c|c} \hline  & Classifier at $R$ & Generator at $A_T$  \\ \hline
$N_R = N_A = 1$   & $0.088$~ms & $0.311$~ms \\ \hline
$N_R = N_A = 4$ & $0.265$~ms & $0.758$~ms \\ \hline
\end{tabular}
\end{table}

\begin{table}
\caption{Inference time of deep neural networks of the receiver and the adversary on FPGA.}
\label{table:fpga}
\small
\centering
\begin{tabular}{c|c|c} \hline  & Classifier at $R$ & Generator at $A_T$  \\ \hline
$N_R = N_A = 1$  & $5.72~\mu$s & $9.04~\mu$s \\ \hline
$N_R = N_A = 4$ & $17.72~\mu$s & $21.04~\mu$s \\ \hline
\end{tabular}
\end{table}

\section{Conclusion}
\label{sec:conclusion}
We developed a GAN-based spoofing attack, where an adversary generates synthetic signals that cannot be reliably distinguished from real signals by using a deep learning-based classifier. We considered both SISO and MIMO communication systems, where each defender or adversary node may have single or multiple antennas. First, we designed a pre-trained deep learning-based classifier to distinguish signals reliably in case there is no spoofing attack. Then, we considered two baseline spoofing attacks, namely transmitting random signals or replaying real signals that are captured previously. Since these spoofing signals cannot keep  patterns of signals received by the intended receiver in terms of waveform, channel, or radio hardware effects, these baseline spoofing attacks are not successful against a deep learning-based classifier. Hence, we designed a spoofing attack building upon a GAN that is trained over the air by an adversary transmitter transmitting synthetic signals and an adversary receiver distinguishing real and synthetic signals. We showed that the success probability of this GAN-based spoofing attack is very high, holds for different network topologies and when node locations change from training to test time, and further improves when multiple antennas are used at the adversary transmitter. Finally, we presented the spoofing attack implementation on embedded platforms and demonstrated the low latency achieved by embedded GPU and FPGA. As the GAN presents a practical threat against intrusion detection mechanisms based on physical layer authentication, future work should look at defense mechanisms to detect and mitigate these novel spoofing attacks.

\end{document}